\documentstyle[aps,prb,preprint]{revtex}

\begin{document}

\draft
\title{Linear theory of unstable growth on rough surfaces}
\author{J.\ Krug$^1$ and  M.\ Rost$^2$}
\address{
$^{(1)}$Fachbereich Physik, Universit\"at GH Essen, 45117 Essen, Germany \\
$^{(2)}$Helsinki Institute of Physics, P.L.\ 9, 00014 University of Helsinki,
Finland\\
}
\date{\today}
\maketitle

\begin{abstract}
Unstable homoepitaxy on rough substrates is treated within
a linear continuum theory. The time dependence of the surface width $W(t)$
is governed by three length scales: The characteristic scale
$l_0$ of the substrate roughness, the terrace size $l_D$ and the
Ehrlich-Schwoebel length $l_{ES}$. If $l_{ES} \ll l_D$ (weak step edge
barriers)
and $l_0 \ll l_m \sim l_D \sqrt{l_D/l_{ES}}$, then $W(t)$ displays a minimum
at a coverage $\theta_{\rm min} \sim (l_D/l_{ES})^2$, where the 
initial surface width is reduced by a factor $l_0/l_m$.
The r\^{o}le of deposition
and diffusion noise is analyzed. The results are applied to recent 
experiments on the growth of InAs buffer layers 
[M.F. Gyure {\em et al.}, Phys. Rev. Lett. {\bf 81}, 4931 (1998)].
The overall features of the observed roughness evolution are captured
by the linear theory, but the detailed time dependence shows distinct
deviations which suggest a significant influence of nonlinearities.

\end{abstract}

\pacs{68.55.Jk, 81.15.Aa, 68.35.Bs, 81.15.Hi}

\section{Introduction}

A high symmetry crystal surface growing epitaxially 
from a molecular beam can become
unstable towards the formation of pyramidal mounds if the
mass transport between different atomic layers is reduced by
additional energy barriers at step edges\cite{villain,note:step_current}. 
Over the last few years, this phenomenon has been observed for a wide range
of metal and semiconductor surfaces, and a considerable body of
theoretical work has been devoted to the description of the
asymptotic (late time) evolution of the surface 
morphology\cite{advances,paris,politirev}.
In the early time regime continuum theory predicts an exponential
growth of the surface modulations. For this reason the precise
initial state of the surface has commonly been disregarded,
since the exponential instability should rapidly
wash out the details of the substrate roughness.

In a recent paper\cite{gyure} Gyure, Zinck, Ratsch and Vvedensky 
(GZRV) presented experimental
and numerical results for the early time development of unstable
homoepitaxy from a rough substrate, which show a more complex
scenario: It was observed that the competition between
smoothening of the initial roughness and the instability
associated with the incipient mound structure can lead to
a {\em minimum} in the total surface width.
A similar effect was predicted previously in the context
of {\em noise-induced} roughening\cite{tapio},
and related experimental observations have been reported 
both for thin metal films\cite{koenig} 
and semiconductor multilayers\cite{klemradt}.
Qualitatively, the minimum originates from
the wavelength dependence of smoothing and
(deterministic or stochastic) roughening rates: If the roughness spectrum
of the substrate has sufficient weight at short wavelengths, which are
efficiently smoothened by capillarity effects\cite{mullins}, then
the decrease of the substrate contribution to the
surface width can temporarily dominate the
long wavelength roughening induced by growth.

The possibility to minimize the surface roughness by an appropriate
choice of the buffer layer thickness and other growth parameters
is of obvious interest in applications. In this paper we
develop a quantitative theory of unstable growth on rough
substrates, which allows us to determine the conditions under
which a minimum occurs, and to 
estimate the layer thickness of minimal roughness in terms
of microscopic length scales and parameters, such as the in-layer
and interlayer diffusion barriers. Our starting point is the
observation of GZRV that
the time evolution of the roughness spectrum appears to be
well described by 
the linearized continuum evolution equation for the surface. 
By incorporating 
various kinds of noise\cite{wolf} into the linear
theory we can compare the influence of stochastic and deterministic
roughening, and obtain a unified description of both cases.
A critical
discussion of our results in relation to the experiments of GZRV
will be presented at the end of the paper. 

\section{Linearized continuum theory}

The standard phenomenological
evolution equation for the continuous surface profile
$H({\bf r},t)$ is of the form\cite{villain,advances,paris,politirev}
\begin{equation}
\label{H}
\partial_t H + \nabla \cdot {\bf J} = F
\end{equation}
where the surface current ${\bf J}$ incorporates both a growth-induced
destabilizing contribution\cite{advances,paris,politirev,politi}
and a stabilizing term originating in capillarity\cite{mullins},
and $F$ denotes
the deposition flux, which will be assumed constant for the time being.
Small fluctuations $h({\bf r},t)$ around the flat singular surface
$H = Ft$ then satisfy the linear equation
\begin{equation}
\label{linear}
\partial_t h = - \alpha \nabla^2 h - \kappa (\nabla^2)^2 h
\end{equation}
with positive coefficients $\alpha$, $\kappa$ representing 
deposition ($\alpha$) and smoothening ($\kappa$), respectively,
whose relation to the
growth parameters will be explained below.

The substrate roughness is incorporated through a spatial roughness
spectrum $\langle \vert \hat h({\bf k},0) \vert^2 \rangle = 
S(k,0) \equiv S_0(k)$,
where $\hat h({\bf k},t)$ is the Fourier transform of $h({\bf r},t)$
and $k = \vert {\bf k} \vert$. Under the linear equation (\ref{linear})
the roughness spectrum evolves as 
\begin{equation}
\label{S(k)}
S(k,t) = S_0(k) \exp[2 (\alpha k^2 -
\kappa k^4) t],
\end{equation}
which implies that fluctuations with
wavenumbers $k > k_c = \sqrt{\alpha/\kappa}$ are damped while those
with $k < k_c$ are exponentially amplified. The surface
width $W(t)$ is obtained by summing over all wavenumbers,
\begin{equation}
\label{W}
W^2(t) = 2 \pi \int_0^\infty dk \; k S_0(k) e^{2(\alpha k^2 - \kappa k^4)t}.
\end{equation}
Motivated by the
experimental data shown in Figure 4 of GZRV
we choose a white noise roughness spectrum,
\begin{equation}
\label{S0}
S_0(k) = \left\{ \begin{array}{l@{\quad:\quad}l}
l_0^2 W_0^2/\pi^3 & k < \pi/l_0 \\
0 & \mbox{else}, \end{array} \right.
\end{equation}
where the small scale cutoff $l_0$ is required for
a finite value $W_0 = W(0)$ of the initial surface width.
Taking the time derivative of (\ref{W}) and evaluating it at
$t=0$, we find that
the surface width shows an initial {\em decrease}
if $(l_m/l_0)^2 > 12$, where $l_m = 2 \pi \sqrt{2 \kappa/\alpha}$
is the wavelength\cite{gyure} of those fluctuations which are maximally amplified
by the linear equation (\ref{linear}). Thus the
condition for a nonmonotonic time dependence of the surface
width is that the length scale characterizing the substrate
roughness, $l_0$, is much smaller than the typical scale
$l_m$ of the emerging mounds. This result holds also for more
general initial roughness spectra, e.g. $S_0(k) = A k^{-\rho}$ with
$\rho < 2$ and a small scale cutoff $l_0$. For substrates
whose roughness is dominated by long wavelength fluctuations, in the
sense that $S_0 \sim k^{-\rho}$ with $\rho > 2$, a large scale cutoff is
needed and the time derivative $dW/dt \vert_{t=0}$ does
usually not exist.

In the following we take $l_0 \ll l_m$. Then
Eq.(\ref{W}) reduces to the scaling form
\begin{equation}
\label{scaling}
W^2(t) = W_0^2(l_0/l_m)^2 \Phi(t/\tau),
\end{equation}
where $1/\tau = \alpha^2/4 \kappa$ is the amplification rate of
the maximally unstable fluctuations and the scaling function
is
\begin{equation}
\label{F}
\Phi(x) = e^{2x}\sqrt{2\pi/x}[1 + {\rm erf}(\sqrt{2x})]
\end{equation}
with ${\rm erf}(s) = (2/\sqrt{\pi}) \int_0^s \exp(-t^2) dt$.
The width attains its minimum at
a time $t_{\rm min} \approx 0.18 \; \tau$, where it has
been reduced by a factor 
\begin{equation}
\label{Wmin}
W(t_{\rm min})/W_0 \approx 3.7 \; (l_0/l_m).
\end{equation}
Since the factor $1 + {\rm erf}(\sqrt{2x})$ in (\ref{F}) only varies
between 1 and 2, the scaling function $\Phi(x)$ is essentially the product
of a decaying power and an exponentially increasing factor. The 
power law for small $x$ reflects the particular smoothening mechanism
(capillarity-driven surface diffusion) and its general form\cite{tapio} 
is given by (\ref{subrough}) below. 
For finite $l_0/l_m$ the power law 
sets in for times $t > t_0$ with
$t_0 \approx (l_0/l_m)^4 \tau$.

To relate the behavior of $W(t)$ to microscopic parameters
we need to express the coefficients $\alpha$ and $\kappa$
of (\ref{linear}) in terms of the two length scales governing
unstable homoepitaxy \cite{advances,politirev}:
The typical terrace size\cite{villainetal} $l_D$ 
and the Ehrlich-Schwoebel-length \cite{politi}
\begin{equation}
\label{schwoebel}
l_{ES} = a_\parallel(D/D' - 1) = a_\parallel (
e^{\Delta E/k_B T}-1)
\end{equation}
defined in terms of the
in-layer lattice spacing $a_\parallel$,
the in-layer (interlayer) surface diffusion constant $D$ ($D'$)
and the step edge barrier $\Delta E$.
Comparison of the two length scales allows to distinguish
conditions of strong ($l_{ES} \gg l_D$) and weak ($l_{ES} \ll l_D$)
step edge barriers; in the first case $\alpha \approx F l_D^2$,
in the second $\alpha \approx F l_D l_{ES}$. The coefficient
$\kappa$ is traditionally associated with near-equilibrium
surface diffusion \cite{mullins}, however under far-from-equilibrium
growth conditions the dominant contribution to $\kappa$ is
believed to arise from the random nucleation process \cite{politi}.
The expression $\kappa \approx F l_D^4$ is then suggested
by dimensional analysis \cite{politi} and scaling arguments \cite{sg}.
It leads to a consistent picture \cite{paris} in the sense that
$l_m \approx l_D$ and $\tau \approx F^{-1}$ in the strong barrier
case, which implies that mounds develop on the submonolayer islands
already during the growth of the first few layers (``wedding cake'' regime
\cite{wed,michely}). In the weak barrier case we find
\begin{equation}
\label{scales}
l_m \sim l_D \sqrt{l_D/l_{ES}} \;\; {\rm and}
\;\;  \tau \sim F^{-1} (l_D/l_{ES})^2.
\end{equation}
The minimum in the surface width thus occurs at a coverage
\begin{equation}
\label{thetamin}
\theta_{\rm min} \sim (l_D/l_{ES})^2 \gg 1
\end{equation} 
which corresponds,
not surprisingly, to the coverage where mounds first become
visible for growth from a smooth substrate \cite{advances,politi}.
Similarly the coverage $\theta_0 = F t_0$ 
at which the scaling form (\ref{scaling}) for the width begins to
hold is of the order
of $\theta_0 \sim (\l_0/l_D)^4$
independent of $l_{ES}$ (provided $l_{ES} \ll l_D$). 

To apply these considerations to the experiment
on InAs growth of GZRV, we first
need to check the condition $l_0 \ll l_m$.
>From Figure 4 of the paper\cite{gyure}
we estimate that $W_0/W(t_{\rm min}) \approx 4$.
Comparing this to the theoretical prediction (\ref{Wmin}) we find
$l_0 \approx 0.07 \times l_m$ and
$l_m/l_0 \gg 1$ is true.
This is in contrast to the kinetic Monte Carlo simulations of
GZRV, where $W_0/W(t_{\rm min}) \approx 1.1$.
The instability length in the
experiment is $l_m \approx 1.0\,\mu$m, which yields
$l_0 \approx 70$ nm for the small scale cutoff of the substrate
roughness. This is consistent with the initial roughness spectrum
in Figure 4 of GZRV, which is constant at least down
to a length scale of 300 nm. The minimum width is attained
at a film thickness of about 0.57 $\mu$m.
Using
$a_\parallel \approx 6 {\rm \AA}$ and a bilayer thickness
$a_\perp \approx 3 {\rm \AA}$, we
therefore estimate that $\theta_{\rm min} \approx 1900$ and
$l_m/a_\parallel \approx 1700$,
and hence $l_{ES}/a_\parallel \approx 6$ and $l_D/a_\parallel
\approx 250$. At the experimental temperature of $500^{\rm o}$ C this
implies a  step edge barrier $\Delta E$
of the order of 0.1 eV, comparable to
estimates\cite{advances,pavel} for GaAs.

\section{Noise effects}

Next we include a noise term $\eta({\bf r},t)$ in Eq.\ (\ref{linear}). 
The different sources of noise, the
individual events of deposition (``shot noise'') and diffusion, enter
the noise correlator with different dependence\cite{wolf,nucleation} 
on the wavenumber $k$. We write it in the form
\begin{equation}
\label{noisecorr}
R(k) \equiv \langle \eta({\bf k},t)
\eta(-{\bf k},t) \rangle = R_{\rm S} + R_{\rm D} k^2 
\end{equation}
with $R_S$ and $R_D$ denoting the strength of deposition and diffusion
noise, respectively. In
the linear model with noise the roughness spectrum $S(k,t)$ then contains a
part reflecting the history of the noise, as well as the 
deterministic evolution of the initial roughness treated above.
The full expression reads
\begin{eqnarray}
\label{total_S}
& S(k,t) & \; \; = \; \; S_{\rm det}(k,t) + S_{\rm noise}(k,t) = \\
& S_0(k) & e^{2(\alpha k^2 - \kappa k^4)t} + \frac{R(k)}{2(\alpha k^2
- \kappa k^4)} \left[e^{2(\alpha k^2 - \kappa k^4)t} - 1
\right]. \nonumber
\end{eqnarray}
Unlike the deterministic mechanism, the noise increases the amplitude
of the spectrum for every wavelength, i.e. $\partial_t
S_{\rm noise}(k,t) > 0$ for all $k$. We shall now examine whether
under the experimental conditions of GZRV noise substantially
contributes to the surface width.

The deposited particle flux can be seen as a Poisson process with
intensity $F$, so $R_{\rm S} \! = \! a_\perp a_\parallel^2 F$. 
Shot noise thus contributes
to the total width by 
\begin{equation}
\label{Ws}
W_{\rm S}(t)^2 = F \tau a_\perp (a_\parallel/l_m)^2
\Psi(t/\tau),
\end{equation}
with $\Psi'(x) = \Phi(x)$ for the choice (\ref{S0}) of
$S_0(k)$. 
Using (\ref{scales}) we see that (\ref{Ws}) can be 
can be ignored against Eq.\ (\ref{scaling}) if
\begin{equation}
\label{condition}
(W_0/a_\perp)^2 (l_0/a_\parallel)^2 \gg (l_D/l_{ES})^2.
\end{equation} 
With our
estimate $l_0 \approx 70$ nm this condition is satisfied in the
experiment. A different interpretation of Eq.(\ref{condition}) 
will be given below. 

The diffusion noise strength is given by the average rate of adatom
jumps on the surface \cite{wolf}, so 
\begin{equation}
R_{\rm D} \! \sim \! \rho_1 D \! \sim \! l_D^2 F
\end{equation}
where we have used the estimate $\rho_1 \sim F l_D^2/D$ for the
adatom density \cite{villainetal}. 
Diffusion noise thus becomes more important than shot noise for $k \!
> \! \pi/l_D$, whereas it can be neglected for long wavelengths. For
large $k$ we can approximate the contribution of diffusion noise in
(\ref{total_S}) by $R_{\rm D} k^2/(\kappa k^4)$ which enters the total
width as $W_{\rm D}(t)^2 = l_D^2/l_m^4 F \tau \log(l_D/a_\parallel)$.
This means that roughly $W_{\rm D}(t)^2 \approx (l_D/l_m)^2 W_{\rm
S}(\tau)^2 \ll W_{\rm S}(\tau)^2$ because
$l_m \gg l_D$ in the weak barrier regime. 
In particular, at the time when the width minimum is
attained, diffusion noise can be neglected against shot noise and for
the experiment of GZRV Eq.\ (\ref{scaling}) remains valid.

It was mentioned already that
due to noise a minimum in the surface width may
occur even in the absence of step edge barriers\cite{tapio}. 
For completeness we provide here
a simple analysis for
the most general linear
Langevin equation of kinetic roughening,
\begin{equation}
\label{Langevin}
\partial_t h = - (- \nu \nabla^2)^{z/2} h + \eta
\end{equation}
where $z=2$ and $z=4$ correspond to evaporation-condensation and
surface diffusion dominated relaxation, respectively \cite{mullins},
$\nu > 0$ is a constant and
and $\eta$ is the deposition noise. Odd or noninteger values of $z$
describe nonlocal relaxation mechanisms and can be treated on the same
footing \cite{advances}. The linearity of (\ref{Langevin}) implies
that the substrate contribution and the growth induced contribution to
the roughness can be separated \cite{tapio}. The substrate contribution
is found to decay according to
\begin{equation}
\label{subrough}
W_{\rm sub}(t) \approx W_0 (l_0/\xi(t))^{d/2}
\end{equation}
for a $d$-dimensional surface, for times such that the correlation
length of the growth-induced roughness $\xi(t)$ exceeds $l_0$
(otherwise $W_{\rm sub} \approx W_0$). In terms of physical quantities the
correlation length can be written as \cite{sg} $\xi(t) \approx l_D
\theta^{1/z}$. To determine the coverage 
$\hat \theta_{\rm min}$ of minimal
surface width for purely stochastic roughening, 
the substrate contribution (\ref{subrough}) should be
compared to the growth induced roughness \cite{sg}
\begin{equation}
\label{growthrough}
W_{\rm growth} \approx a_\perp (\theta/\tilde \theta)^{(z-d)/2z}
\end{equation}
where $\tilde \theta$ is the coverage at which the width becomes of
the order of $a_\perp$ and thus lattice effects (such as temporal
oscillations of the step density) die out; the expression
(\ref{growthrough}) holds for $\theta \gg \tilde \theta$. With the
estimate \cite{sg,kbkw} $\tilde \theta \sim
(l_D/a_\parallel)^{zd/(z-d)}$ we obtain
\begin{equation}
\label{minrough}
\hat \theta_{\rm min} \approx (W_0/a_\perp)^2 (l_0/a_\parallel)^d
\end{equation}
independent of $z$. 
Comparing Eqs. (\ref{minrough}) and (\ref{thetamin}) we have thus recovered
the crossover condition (\ref{condition})
between deterministic instability and stochastic roughening
from the opposite side. To neglect the growth
instability in Eq.(\ref{Langevin}) is no longer justified when the minimum 
width coverage $\theta_{\rm min}$ 
predicted by the deterministic theory (Eq.(\ref{thetamin})) is
smaller than $\hat \theta_{\rm min}$ in Eq.(\ref{minrough}).

\section{Discussion and conclusion}

The main results of this paper are Eqs. (\ref{scaling}) and 
(\ref{Ws}), which express the time dependent surface roughness
in terms of the characteristic length and time scales of the problem -- 
the substrate roughness scale $l_0$, the
incipient mound size $l_m$, and
the linear growth time $\tau$ --, the latter two of which are, in turn,
related to the microscopic growth parameters through Eq.(\ref{scales}). 
For the experiment of GZRV, the measured values of $l_m$
and $\tau$ were seen to imply reasonable numbers for the microscopic
lengths $l_D$ and $l_{ES}$, and for the step edge barrier $\Delta E$. 

It is then natural to ask to what extent the linear theory can be
used to {\em quantitatively} describe the 
experimentally observed roughness evolution, beyond providing
consistent order-of-magnitude estimates. Inspection of the
data for the surface roughness 
depicted in the inset of Figure 4 of GZRV 
quickly leads to the conclusion that, despite a similar overall appearance,
the shape of $W(t)$ is {\em not} well reproduced by 
our scaling functions (\ref{scaling}) and (\ref{Ws}). 
In fact the data for the structure factor in Figure 4
show a {\em qualitative} feature which the linear theory is unable to
explain: It is an immediate consequence of Eq.(\ref{total_S}) that 
$S(k,t)$ is a monotonic function of $t$ (increasing or decreasing) for 
any $k$; in contrast, the measured structure factor
shows a nonmonotonic dependence on film thickness for $k > k_c$. 

This prompts the question whether the use of the linearized theory
is really justified under the experimental conditions. 
Nonlinear terms in the surface current ${\bf J}$ in Eq.(\ref{H})
are expected\cite{advances,paris,politi} to matter when 
the surface slope $\vert \nabla h \vert$ becomes comparable to
$a_\perp/l_D$. Since typical slope values of
the initial surface profile are of the order of $W_0/l_0$, the condition
for the validity of the linear theory is 
\begin{equation}
\label{lincond}
W_0/a_\perp < l_0/l_D.
\end{equation} 
In the experiment of 
GZRV $W_0/a_\perp \approx 3$ and, from the estimates
presented above, $l_0/l_D \approx 0.5$; thus the condition
(\ref{lincond}) is (weakly) violated.  
The analogy to phase ordering kinetics\cite{paris} 
suggests that the early time evolution may be qualitatively
altered when nonlinearities are important\cite{claudio}.
This seems like an interesting problem for future study.  

\vspace{0.5cm}

\noindent
{\em Acknowledgements.} We thank P. \v{S}milauer and M. Siegert
for useful discussions, and M. Gyure for providing us with
experimental data. This work was supported by DFG within SFB 237
and by the European Union within TMR network Nr.\ ERB 4061 PL 97-0775.


\begin{references}
\bibitem{villain} J. Villain, J. Phys. I France {\bf 1}, 19 (1991).
\bibitem{note:step_current} We assume here that the mounding instability
is caused by step edge barriers suppressing {\em interlayer} transport. 
The possibility of {\em corner} barriers which bias the 
{\em in-layer} diffusion of atoms along step edges causing an instability
has been pointed out recently by several groups 
[O. Pierre-Louis, M.R. D'Orsogna and T.L. Einstein, 
Phys. Rev. Lett. {\bf 82}, 3661 (1999); M.V. Ramana Murty and B.H. Cooper,
Phys. Rev. Lett. {\bf 83}, 352 (1999)], however
for this scenario the continuum theory is yet to be worked out 
[P. Politi and J. Krug, preprint (cond-mat/9908152)]. 
\bibitem{advances} J. Krug, Adv. Phys.
{\bf 46}, 139 (1997).
\bibitem{paris} J. Krug, Physica A {\bf 263}, 170 (1999).
\bibitem{politirev}
P. Politi, G. Grenet, A. Marty, A. Ponchet and J. Villain,
Phys. Rep. (in press).
\bibitem{gyure} M.F. Gyure, J.J. Zinck, C. Ratsch and 
D.D. Vvedensky, Phys. Rev. Lett. {\bf 81},
4931 (1998).
\bibitem{tapio} S. Majaniemi, T. Ala-Nissila and 
J. Krug, Phys. Rev. B {\bf 53},
8071 (1996).
\bibitem{koenig} F. K\"onig, PhD thesis, KFA J\"ulich Report No.3092 (1995).
\bibitem{klemradt} U. Klemradt, M. Funke, M. Fromm, B. Lengler, J. Peisl
and A. F\"orster, Physica B {\bf 221}, 27 (1996).
\bibitem{mullins} W.W. Mullins, J. Appl. Phys. {\bf 30}, 77 (1959).
\bibitem{wolf} D.E. Wolf, in ``Scale Invariance, Interfaces, and
Non-Equilibrium Dynamics'', ed. by A. McKane, M. Droz, J. Vannimenus
and D. Wolf (Plenum Press, New York 1995), p. 215.
\bibitem{politi} P. Politi and J. Villain, Phys. Rev. B {\bf 54},
5114 (1996).
\bibitem{villainetal} J. Villain, A. Pimpinelli, L. Tang and D. Wolf,
J. Phys. I France {\bf 2}, 2107 (1992).
\bibitem{sg} M. Rost and J. Krug, J. Phys. I France {\bf 7},
1627 (1997).
\bibitem{wed} J. Krug, J. Stat. Phys. {\bf 87}, 505 (1997).
\bibitem{michely} M. Kalff, P. \v{S}milauer, G. Comsa and Th. Michely,
Surf. Sci. Lett. {\bf 426}, L447 (1999).
\bibitem{pavel} P. \v{S}milauer and D.D. Vvedensky, Phys. Rev. B {\bf 48},
17603 (1993).
\bibitem{nucleation} Nucleation also consists of random events and
enters $R(k)$ with a term\cite{wolf} $R_{\rm N} k^4$. 
However it is dominated by the
other noise sources for all $k$ up to
$\pi/a_\parallel$.
\bibitem{kbkw} H. Kallabis, L. Brendel, J. Krug and
D.E. Wolf, Int. J. Mod. Phys. B {\bf 11}, 3621 (1997).
\bibitem{claudio} C. Castellano and M. Zannetti, Phys. Rev. E {\bf 58},
5410 (1998).


\end{references}
\end{document}